\documentclass[aps,pra,twocolumn,showpacs]{revtex4-1}
\usepackage{graphicx,amsmath}
\newcommand{\bra}[1]{\langle{#1}\vert}
\newcommand{\ket}[1]{\vert{#1}\rangle}
\begin{document}
\title{Weak measurement-based state estimation 
of Gaussian states of one-variable quantum systems}
\author{Debmalya Das}
\email{debmalya@iisermohali.ac.in}
\affiliation{Department of Physical Sciences,
Indian Institute of Science Education \& Research
(IISER) Mohali, Sector-81, SAS Nagar, Manauli P.O.
140306, Punjab, India.}
\author{Arvind}
\email{arvind@iisermohali.ac.in}
\affiliation{Department of Physical Sciences,
Indian Institute of Science Education \& Research
(IISER) Mohali, Sector-81, SAS Nagar, Manauli P.O.
140306, Punjab, India.}
\begin{abstract}
We present a scheme to estimate Gaussian states of
one-dimensional continuous variable systems, based
on weak (unsharp) quantum measurements.  The
estimation of a Gaussian state requires us to find
position ($q$), momentum ($p$) and their second
order moments.  We measure $q$ weakly and follow
it up with a projective measurement of $p$ on half
of the ensemble, and on the other half we measure
$p$ weakly followed by a projective measurement of
$q$. In each case we use the state twice before
discarding it.  We compare our results with
projective measurements and demonstrate that under
certain conditions such weak measurement-based
estimation schemes, where recycling of the states
is possible, can outperform projective
measurement-based state estimation schemes. 
\end{abstract} 
\maketitle
\section{Introduction} 
\label{introduction}
State determination for a physical system hinges on being
able to non-invasively measure its relevant parameters.
However, a measurement performed on a quantum system is by
definition invasive, and hence quantum state estimation
relies on measurements made on an ensemble of identically
prepared systems. Quantum state estimation process is thus
intrinsically statistical in nature, with its accompanying
ambiguities and uncertainties.~\cite{sakurai, griffiths}.
There is no direct measurement possible for the quantum
state of a single system and for the estimation of a state,
we are required to determine the expectation values of a set
of incompatible observables. The accuracy of such a
determination  depends upon the size of the ensemble and
ideally we need an infinite size ensemble to obtain the
precise values of these expectations.  However, if we are
given a small ensemble with a fixed number of identically
prepared states, how can we effectively extract information
from it? The wave function collapse associated with
projective measurements limits us from using each
member of
the ensemble more than once.  The problem of state
estimation has been investigated by physicists since the
inception of quantum information theory~\cite{NC}.  Apart
from the direct use of  projective measurements, there exist
other methods of state tomography which try to extract
information from a system in different ways. Some
prescriptions use the information gained about the system
from one measurement to decide on the next
measurement~\cite{adapt}.  Others
employ maximum likelihood technique and
numerical optimization~\cite{Daniel2001}, or use
repeated weak measurements on a single
system~\cite{Diosi}, to maximize the information
gain. Weak or unsharp measurements potentially
hold promise for state estimation because of their
non-invasiveness which allows state recycling. On
the one hand the disturbance caused by a weak
measurement is less, however on the other hand it
also gives us limited information.  The challenge
therefore is to find a balance i.e. an
intermediate regime of weakness, which leads to
optimal information gain.  The idea has been
explored recently in the context of a qubit with a
small ensemble size~\cite{Qsize,Arvind2015}.  Weak
measurements coupled with postselection have also
been employed in the problem of state estimation
where a  complete characterization of the
postselected quantum statistics~\cite{Hofmann} or
the direct measurement of the quantum wavefunction
is used~\cite{Qwavefn,GenQwavefn, phyQwavefn}.
State measurement schemes based on weak
measurement tomography have also been recently
proposed~\cite{sttomowm,sswm}. There have also
been critial analysis of these
schemes~\cite{gross}.

For continuous variable systems, quasi-probability
distributions including the Wigner distributions can be
tomographed by measuring the rotated quadrature
components~\cite{Vogel}. There are homodyne and heterodyne
schemes for estimation of squeezed Gaussian
states~\cite{Rehacek2015} and squeezed thermal Gaussian
states~\cite{Aspachs2009,Welsch2009}.  Schemes using a
single photon detector instead of a homodyne detector to
characterize Gaussian states have been
proposed~\cite{Wenger2004}.  The possible advantage offered
by  an entangled Gaussian probe to estimate the displacement
of a continuous variable state has also been
explored~\cite{Adesso2013}.  The importance of maximum
likelihood methods has been emphasized~\cite{Hadril1997,
Banaszek1999} and state reconstruction has been
described~\cite{Lvovsky2009, Mallet2011}.  In a different
direction, Arthurs and Kelly aimed to simultaneously measure
position  and momentum of a general wavefunction by coupling
two different apparatuses with the system~\cite{kelly}.
Since $q$ and $p$ are noncommuting observables, 
this leads to an unsharp measurement. Symplectic
tomography has been used to estimate the master
equation parameters in an open setting for a single
mode system~\cite{ugo}.

Alternatively, one can reuse each member of the
ensemble if the first measurement is done weakly
enough such that the disturbance induced is very
small~\cite{brun, Qsize}. A similar idea has been
utilized, albeit in a different direction, in the
construction of loophole free hybrid Bell-Leggett-
Garg inequalities~\cite{Dressel2011, Dressel2014}.
Weak or unsharp measurements are performed by
weakly coupling the device to the quantum
system~\cite{AAV, Sudweak, Joz, Qsize}. Although
the noise produced in such measurements is small,
which should serve our purpose well, the
information obtained is also very low.  Therefore,
there is a tradeoff between the disturbance and
information gain. To effectively use weak
measurements for state estimation, we need to
optimize the process.

In this work we restrict ourselves to the  realm
of a special class of states of one continuous
variable quantum systems called Gaussian
states.  These  are states with Gaussian-Wigner
quasiprobability distributions and include
coherent states, squeezed states and thermal
squeezed states. The Gaussian states are
determined by the first and second moments of
position and momentum.  We explore the
advantage of the scheme involving weak
measurements in estimating the Gaussian states
over projective measurements. We show that with
an optimal strength of the weak measurement, our
technique is more powerful in the determination of
the Wigner quasiprobability distribution when the
ensemble size available is small. We have
chosen the meter state in the form of a minimum
uncertainty squeezed Gaussian state and the tuning
of the weakness of a measurement is achieved by
changing the squeezing of the position quadrature.
This scheme is tested for average performance over
a large number of states and over a large number
of runs to kill statistical fluctuations. We also
take Gaussian states at different temperatures and
check whether the efficacy of our method depends
in any way on temperature.

The paper is arranged as follows:  In
section~\ref{tools} we collate all the background
material necessary for the problem. 
We
describe continuous variable states and Wigner's
quasiprobability distributions in brief.
Symplectic methods are described briefly.
Section~\ref{scheme}  gives a description of weak
measurement in quantum mechanics when applied to
Gaussian states. In section~\ref{tomo} we describe
how to perform tomography of Gaussian states using our
method.
We provide conclusions in section~\ref{conc}.
\section{Gaussian States}
\label{tools}
Let us consider a one-dimensional quantum system
with the position and momentum operators $\hat{q}$ and 
$\hat{p}$, satisfying the commutation relation
\begin{equation}
[\hat{q},\hat{p}]=\iota \quad \rm{(}\hbar=1\rm{)}
\label{commutation}
\end{equation}
The corresponding eigenkets are $\ket{q}$ and
$\ket{p}$ for a complete set~\cite{sakurai}, and are defined as
\begin{eqnarray}
&\hat{q}\ket{q}=q\ket{q}\rm{,} \;\;
\hat{p}\ket{p}=p\ket{p} &\nonumber \\
&\!\!\!\!\langle q \vert
q^\prime\rangle=\delta(q-q^\prime), \;
\langle p \vert p^\prime\rangle =
\delta(p-p^\prime)\, {\rm and}\,
\langle q\vert p\rangle =e^{\iota pq}&
\end{eqnarray}
An arbitrary mixed density operator in position and
momentum bases can be represented by
\begin{eqnarray} \hat{\rho}(q,q^\prime)=\int dq
dq^\prime \;
f(q,q^\prime)\ket{q}\bra{q^\prime}\nonumber\\
\hat{\tilde{\rho}}(p,p^\prime)=\int dp dp^\prime
\; \tilde{f}(p,p^\prime)\ket{p}\bra{p^\prime}
\label{docontinuous} 
\end{eqnarray} 
where
$f(q,q^\prime)$ is a function of 
real variables $q$ and $q^\prime$ while
$\tilde{f}(p,p^\prime)$ is a function of the real
variables $p$ and $p^\prime$.

An alternative and equivalent way of representing the
system state~\cite{Wigner1932} is via its Wigner distribution
$W(q,p)$ given by \begin{equation}
W(q,p)=\frac{1}{2\pi}\int_{-\infty}^\infty dy\;
\bra{q-\frac{y}{2}}\hat{\rho}\ket{q-\frac{y}{2}}
e^{\iota py} \label{wigdef} 
\end{equation}
The probability distributions corresponding
to position and momentum can be obtained by computing
the marginals
\begin{eqnarray}
P(q)&=& \int^\infty_{-\infty} W(q,p) dp \nonumber
\\
P(p)&=& \int^\infty_{-\infty} W(q,p) dq
\end{eqnarray}
and can be used to calculate expectation values of
arbitrary observables via the symmetric ordering
rule~\cite{Wigner1932, nmode}.

The second order moments of position and momentum
corresponding to a quantum state are given by 
\begin{eqnarray}
(\Delta q)^2 &=& \langle q^2\rangle-\langle q\rangle^2 \rm{,}\;
(\Delta p)^2=\langle p^2\rangle-\langle p\rangle^2\nonumber\\
\Delta (q,p) &=& \frac{1}{2}\langle \{\hat{q}-\langle \hat{q}\rangle,
\hat{p}-\langle \hat{p}\rangle\}\rangle
\end{eqnarray} 
and they obey the 
Schr\"odinger uncertainty principle  given by
\begin{equation}
(\Delta q)^2 (\Delta p)^2\geq \vert \frac{1}{2}
\langle\{\hat{q},\hat{p}\}\rangle-
\langle\hat{q}\rangle\langle\hat{p}\rangle\vert^2
+\vert\frac{1}{2\iota}\langle[\hat{q},\hat{p}]\rangle\vert^2
\end{equation}
A compact way to represent the second order
moments is via the variance matrix $V$ given by
\begin{equation}
 \label{gen variance matrix}
  V=\begin{pmatrix}
     (\Delta q)^2 & \Delta (q, p)\\
     \Delta (q, p) & (\Delta p)^2
    \end{pmatrix}
 \end{equation}
and the uncertainty condition re-expressed in
terms of the  
variance matrix takes the elegant form~\cite{nmode, mukundabell}
\begin{eqnarray}
V+\frac{\iota}{2}\beta\equiv \rm{positive\;
semidefinite} \nonumber \\
\beta=\begin{pmatrix}
 0 & 1\\
-1 & 0
\end{pmatrix}, \,
\beta^{-1}=\beta^T
=-\beta, \,{\rm Det}(\beta)=1
\label{wignerur}
\end{eqnarray}
The subclass of states for which the Wigner
distribution is a Gaussian function  are called
Gaussian states and play an important role in
quantum optics and quantum information
~\cite{Olivares2012, squeeze, pedestrian}.  All
states with Gaussian wave functions are Gaussian,
however the class of Gaussian states is a much bigger
class and includes mixed states.  The Wigner
representation corresponding to a general Gaussian
state centered at the origin of the phase space is
given by,
\begin{equation}
W\left(\xi\right)=\frac{1}{\pi}\sqrt{|G|}e^{-\xi^T G \xi}
\label{Gaucen}
\end{equation}
where
\begin{eqnarray}
 \xi &=& \begin{pmatrix}
            q\\  p
           \end{pmatrix} \nonumber \\
 G &=&G^* =G^T
 \end{eqnarray}
The matrix $G$ is related to the variance matrix $V$ as 
\begin{equation}
 V=\frac{1}{2}G^{-1}
\end{equation}
If the center of the Gaussian is located at a point $(q_0,p_0)$, this can be 
achieved by action of a displacement operator $\hat{D}(q_0,p_0)$ which
acts on canonical operators as 
\begin{eqnarray}
&\hat{\xi} =\begin{pmatrix}
\hat{q} \\ \hat{p}
\end{pmatrix} 
\rightarrow \hat{D}(q_0,p_0)\hat{\xi}\hat{D}(q_0,p_0)^{-1}=
\begin{pmatrix}
\hat{q}-q_0 \\ \hat{p}-p_0
\end{pmatrix}
=\hat{\tilde{\xi}}& \nonumber \\
&\hat{D}(q_0,p_0)=e^{\iota\left(p_0
\hat{q}-q_0\hat{p}\right)}&
\label{displacement}
\end{eqnarray}
and leads to a point transformation of the Winger
function. For the Gaussian-Wigner function this
amounts to shifting the center to the location
$(q_0,p_0)$.
The matrix $G$ can be written as:
\begin{equation}
 G=\hat{U}S^T G_0 S\hat{U}^{-1}
\label{theG}
\end{equation}
where $S$ is a diagonal symplectic matrix
belonging to the group $Sp(2,R)$, $\hat{U}$ is a rotation 
matrix and $G_0$ is proportional to identity. A
total of three real parameters are involved in
describing $G$. We further restrict ourselves to
a special class of Gaussians where $G$ is described
by two parameters, namely, the temperature and
squeezing~\cite{nmode, mukundabell}. Setting  
\begin{equation}
\hat{U}=I, \quad
S = \begin{pmatrix}
      e^{-u} && 0\\
      0 && e^u
\end{pmatrix}\rm{,}\quad
G_0 = \kappa I
\label{squeezing}
\end{equation}
with
$\kappa = \tanh \left(\frac{ \omega}{2kT}\right)$
representing the temperature $T$,  
$k$ being the Boltzmann's constant and $\omega$
having the units of frequency.

\begin{figure}[t]
\centering
\includegraphics[scale=1]{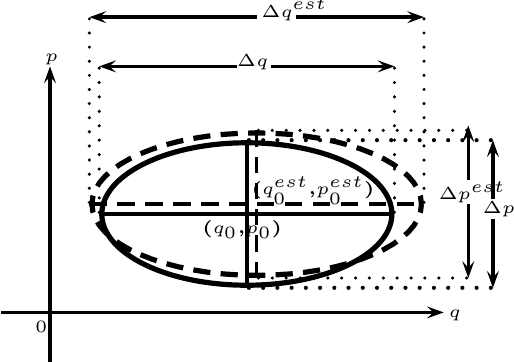}
\caption{The Gaussian state, represented by an
ellipse, in the phase space. The actual state is
an ellipse bounded by continuous line, centered at
$\left(q_0,p_0\right)$ and with spreads $\Delta q$
and $\Delta p$.  The estimated state is
represented by another ellipse, bounded by a
broken line, centered at
$\left(q^{est}_0,p^{est}_0\right)$ and has spreads
$\Delta q^{est}$ and $\Delta p^{est}$. 
\label{spread}}
\end{figure}

The corresponding variance matrix in this case is
diagonal 
\begin{equation}
V=\begin{pmatrix}
(\Delta q)^2 & 0\\
0 & (\Delta p)^2
\end{pmatrix}
\label{variance matrix}
\end{equation}
At $T=0 K$ or $\kappa=1$ if we have $\Delta
q=\Delta p=\frac{1}{\sqrt{2}}$, the Gaussian state
is a coherent state. A coherent state is
represented by a circle of radius
$\frac{1}{\sqrt{2}}$ in phase space. Now at $T=0
K$, if $\Delta q$ and $\Delta p$ happen to be
unequal, the state is a  squeezed state. Such a
Gaussian state is represented by an ellipse. The
center of a general coherent or a general squeezed
state may not be at the origin of phase space and
in such cases it is said to be a displaced
coherent or a displaced squeezed state. The displacement of a
Wigner function $W(\xi)$, centered at the origin
is achieved by means of a displacement operator
$\hat{D}(q_0,p_0)$ which takes the center to
$(q_0,p_0)$.  For a non-zero temperature, the
value of the product $\Delta q \Delta p$ is
greater than $\frac{1}{2}$ (for coherent states it
is equal to half) and it
increases with a rise of temperature. Pictorially,
the class of states can be represented by an
ellipse in phase space with center at $(q_0,p_0)$
and semimajor axis oriented along either $q$ or
$p$ and is depicted in Figure~\ref{spread}.
\section{Weak measurements on Gaussian states}
\label{scheme}
Consider a system represented by a displaced
Gaussian-Wigner distribution function represented
by $W_s(\xi_s)$ and characterized by two real
displacement parameters: one temperature parameter
and one squeezing parameter. Such a Wigner
function can be obtained by starting with the
centered Gaussian-Wigner function given in
Equation~(\ref{Gaucen}) with parameters chosen as
per Equations~(\ref{theG})~and~(\ref{squeezing})
and applying a displacement operator $\hat{D}(q_0,p_0)$
as described in the previous section. Given that
the system is in such a state, our goal  is to
estimate the state.

For the purposes of measurement, consider a meter
which is a macroscopic pointer with position and
momentum variables $q_{m_1}$ and $p_{m_1}$.  The
meter is chosen to be in a squeezed coherent state
represented by a Wigner distribution $W(\xi_{m_1})$
such  that $\Delta q_{m_1} \Delta
p_{m_1}=\frac{1}{2}$ and $\Delta q_{m_1}\neq
\Delta p_{m_1}$.  Here we take temperature to be
zero.  As we shall see, when we employ this meter
to measure the position, the strength of the
measurement can be varied by changing the
squeezing of the meter along the position
quadrature. The larger values of the squeezing
parameter correspond to a stronger measurement
while the smaller values of the squeezing
parameter correspond to a weaker measurement.
Similarly, if we are measuring momentum one can
tune the measurement strength by varying the value
of squeezing of the  variable $q_{m_2}$.

The system and the meter form  a composite system
and the joint  Wigner function representing this
two degrees of freedom system can be obtained by
multiplying the two individual Wigner functions.  
For such a system it is natural to define a four
dimensional column vector of
phase space variables as
\begin{equation}
{\Xi}=\begin{pmatrix}
{q}\\
{q}_{m_1}\\
{p}\\
{p}_{m_1}
\end{pmatrix}.
\end{equation}
The phase space displacement of the system
variables given in Equation~(\ref{displacement}) 
acts on this column vector to give us a displaced
vector 
\begin{equation}
{\tilde{\Xi}}=\begin{pmatrix}
{q}-q_0\\
{q}_{m_1}\\
{p}-p_0\\
{p}_{m_1}
\end{pmatrix}.
\end{equation} 
In terms of the above column vector
The joint Wigner function becomes
\begin{eqnarray} 
W\left(\Xi\right) =
\frac{1}{\pi^2}\sqrt{|G|}e^{-\tilde{\Xi}^T G
\tilde{\Xi}} \label{defwig}
\end{eqnarray}
The matrix $G$ is a diagonal $4\times4$ matrix
\begin{equation}
G={\rm Diag} 
\left((\Delta q)^2, 
(\Delta q_{m1})^2,
 (\Delta p)^2 ,
(\Delta p_{m1})^2
\right)
\end{equation}

\begin{equation}
W(\Xi)=\frac{\exp{\left[\scriptstyle{-\frac{1}{2}
\left(\frac{(q-q_0)^2}{(\Delta q)^2}
+\frac{(p-p_0)^2}{(\Delta
p)^2}+\frac{q_{m_1}^2}{(\Delta q_{m_1})^2}
+\frac{p_{m_1}^2}{(\Delta
p_{m_1})^2}\right)}\right]}}{4\pi^2 \Delta q
\Delta p \Delta q_{m_1} \Delta p_{m_1}}
\end{equation}
When we perform a measurement (weak or strong) 
of the position $q$, we switch on the
following interaction Hamiltonian between the 
system and the meter degrees of freedom 
\begin{equation}
\hat{H}=\delta(t-t_1)\hat{q} \hat{p}_{m_1}
\end{equation}
The corresponding unitary transformation on the composite
system-meter is
\begin{equation}
 \hat{U}=e^{-\iota \int \hat{H} dt}
\end{equation}
In the language of the Wigner quasi-probability
distribution a unitary operation $\hat{U}$ is
equivalent to a symplectic transformation
$\mathcal{S}$
\begin{equation}
\mathcal{S}=\begin{pmatrix}
           1 & 0 & 0 & 0\\
           1 & 1 & 0 & 0\\
           0 & 0 & -1 & 1\\
           0 & 0 & 0 & 1
         \end{pmatrix}
\end{equation}
satisfying
\begin{equation}
\mathcal{S}^T\beta_2 
\mathcal{S}=\mathcal{S}\beta_2 
\mathcal{S}^T=\beta_2\,{\rm with}\,
\beta_2=\begin{pmatrix}
0_{2\times2} & I_{2\times2}\\
-I_{2\times2} & 0_{2\times2}
\end{pmatrix}.
\end{equation}
The symplectic transformation corresponding to the
interaction Hamiltonian acts on the phase
variables by multiplication 
\begin{equation}
\Xi^{\prime} = S \Xi
\end{equation}
leading to the computation of the 
transformed  Wigner
function under this symplectic transformation 
\begin{equation}
W^\prime(\Xi)=\frac{\exp{\left[\scriptstyle{-\frac{1}{2}
\left(\frac{(q-q_0)^2}{(\Delta
q)^2}+\frac{(p+p_{m_1}-p_0)^2} {(\Delta p)^2}
+\frac{(q_{m_1}-q)^2}{(\Delta
q_{m_1})^2}+\frac{p_{m_1}^2}{(\Delta
p_{m_1})^2}\right)}\right]}} {4\pi^2 \Delta q
\Delta p \Delta q_{m_1} \Delta p_{m_1}}.
\end{equation} 
The Wigner function of the meter after the above
interaction is obtained by integrating over the
system variables $q$ and $p$ and is given by

\begin{equation}
W^\prime_{m_1}(\xi_{m_1})=
\frac{\exp{\left[\scriptstyle{-\frac{1}{2}
\left(\frac{(q_{m_1}-q_0)^2}{(\Delta q_{m_1})^2
+(\Delta q)^2}+\frac{p_{m_1}^2}{(\Delta p_{m_1})^2}
\right)}\right]}}{2\pi\Delta q_{m_1}
\Delta p_{m_1}\Delta q \sqrt{\frac{1}{(\Delta
q_{m_1})^2}+\frac{1}{(\Delta q)^2}}}.
\label{wigsymp1}
\end{equation}
We can see at this point that the state of the
meter has become correlated with the state of the
system. However, as the meter is a macroscopic
entity, its very observation leads to the collapse
of its wavefunction and gives us a
definite value. 
Thus the probability density for the
meter to show a reading $q_{m_1}$
\begin{equation}
P(q_{m_1})=\frac{\exp{\left[\scriptstyle{-\frac{1}{2}
\left(\frac{(q_{m_1}-q_0)^2}{(\Delta q_{m_1})^2
+(\Delta q)^2}\right)}\right]}} {\sqrt{2\pi}
\Delta q_{m_1} \Delta q\sqrt{\frac{1}{(\Delta
q_{m_1})^2} +\frac{1}{(\Delta q)^2}}}.
\label{probweak1}
\end{equation}
On the other hand, the reduced state of the system
after the measurement interaction represented by
the symplectic transformation is obtained by
integrating over the meter degrees of freedom
leading to the Wigner function for the system 
\begin{equation}
W^\prime_s(\xi_s)=\frac{\exp{\left[\scriptstyle{-\frac{1}{2}
\left(\frac{(p-p_0)^2}{(\Delta p_{m_1})^2+(\Delta
p)^2} +\frac{(q-q_0)^2}{(\Delta
q)^2}\right)}\right]}} {2 \pi \Delta p_{m_1}
\Delta p \Delta q \sqrt{\frac{1}{(\Delta
p_{m_1})^2}+\frac{1}{(\Delta p)^2}}}
\label{wigtrans}
\end{equation}
In the weak measurement limit $\Delta q_{m_1}$ is
large (i.e. the initial meter state is prepared in
distributions wide in position).
Since we have chosen the meter to be in a
squeezed coherent state, this corresponds to a
high degree of squeezing in the momentum
quadrature of the initial meter state.
In this limit we have 
\begin{equation}
\Delta
p_{m_1}\rightarrow 0 \Rightarrow
W^\prime_s\rightarrow
W_s.
\end{equation}
Hence, weak measurement causes controllable
disturbance to the state and the the disturbance
vanishes in the limit of extremely weak measurement.
However, if we make the measurement too weak, the
correlation between the meter state and the system
state diminishes. In the limit of
extremely weak measurement, where no disturbance is
caused, we do not learn anything about the system
from observing the meter.

In our scheme, the first measurement that we
perform is a weak measurement of position $q$ with a
tunable strength as described above.
Subsequently, we carry out a projective measurement
of momentum $p$ is on this system, then the
probability density for obtaining any momentum
as obtained from the modified system Wigner
function given in Equation~(\ref{wigtrans})  is given by,
\begin{equation}
P(p)=\frac{\exp{\left[\scriptstyle{-\frac{1}{2}
\left(\frac{(p-p_0)^2}{(\Delta p_{m_1})^2+ (\Delta
p)^2}\right)}\right]}} {2 \pi \Delta p_{m_1}
\Delta p \sqrt{\frac{1}{(\Delta
p_{m_1})^2}+\frac{1}{(\Delta p)^2}}}
\label{probstrong1}
\end{equation}

In the reverse scenario where we do a weak
measurement of momentum $p$ followed by a projective
measurement of position $q$,
the
composite system-meter system Wigner function
after the measurement interaction given by the
Hamiltonian
\begin{equation}
\hat{H^\prime}=\delta(t-t^{\prime}_1)\hat{p} \hat{p}_{m_2}
\end{equation}
is given by
\begin{equation}
W^{\prime\prime}(\Xi)=\frac{\exp{\left[\scriptstyle{-\frac{1}{2}
\left(\frac{(p-p_0)^2}{(\Delta
p)^2}+\frac{p_{m_2}^2} {(\Delta
p_{m_2})^2}+\frac{(q-p_{m_2}-q_0)^2}{(\Delta
q)^2}+\frac{(p-q_{m_2})^2} {(\Delta
q_{m_2})^2}\right)}\right]}}{4 \pi ^2\Delta
p_{m_2} \Delta p \Delta q_{m_2} \Delta q} 
\label{wigsymp2}
\end{equation}
where $q_{m_2}$ and $p_{m_2}$ denote the position
and momentum coordinates of the meter measuring
momentum $p$ of the system. 
The Wigner 
of the meter alone is given by
\begin{equation}
W^{\prime\prime}_{m_2}(\xi_{m_2})=
\frac{\exp{\left[\scriptstyle{-\frac{(q_{m_2}-p_0)^2}{2
\left((\Delta p)^2 +(\Delta q_{m_2})^2\right)}
-\frac{p_{m_2}^2}{2 (\Delta
p_{m_2})^2}}\right]}}{2 \pi  \Delta p_{m_2} \Delta
p \Delta q_{m_2} \sqrt{\frac{1}{(\Delta
p)^2}+\frac{1}{(\Delta q_{m_2})^2}}}
\end{equation}
giving the  probability density of the meter to show a
reading $q_{m_2}$ being
\begin{equation}
P(q_{m_2})=\frac{\exp{\left[\scriptstyle{-\frac{(q_{m_2}-p_0)^2}{2
\left((\Delta p)^2 +(\Delta
q_{m_2})^2\right)}}\right]}} {\sqrt{2 \pi}  \Delta
p \Delta q_{m_2} \sqrt{\frac{1}{(\Delta
p)^2}+\frac{1}{(\Delta q_{m_2})^2}}}
\label{probweak2}
\end{equation}
The corresponding system Wigner function becomes
\begin{equation}
W^{\prime\prime}_s(\xi_s)=\frac{\exp{\left[\scriptstyle{-\frac{1}{2}
\left(\frac{(p-p_0)^2}{(\Delta p)^2}
+\frac{(q-q_0)^2}{(\Delta p_{m_2})^2 +(\Delta
q)^2}\right)}\right]}}{2 \pi  \Delta p_{m_2}
\Delta p \Delta q \sqrt{\frac{1}{(\Delta
p_{m_2})^2}+\frac{1}{(\Delta q)^2}}}
\end{equation}
As before, in the weak measurement limit the
disturbance caused in the system is limited and we 
have 
\begin{equation}
\Delta
q_{m_2}\rightarrow 0 \Rightarrow
W^{\prime \prime}_s\rightarrow
W_s.
\end{equation}

On 
this state we perform a projective measurement of position
$q$ giving us the  probability density for
getting a result $q$
\begin{equation}
 P(q)=\frac{\exp{\left[\scriptstyle{-\frac{1}{2}\left(
\frac{(q-q_0)^2}{(\Delta p_{m_2})^2 +(\Delta
q)^2}\right)}\right]}}{\sqrt{2 \pi} \Delta p_{m_2}
\Delta q \sqrt{\frac{1}{(\Delta
p_{m_2})^2}+\frac{1}{(\Delta q)^2}}}.
\label{probstrong2}
\end{equation}
\section{Estimation of Gaussian states using weak measurements}
\label{tomo}
\subsection{The prescription}
\label{pres}
\begin{figure}[t]
\centering
\includegraphics[scale=1]{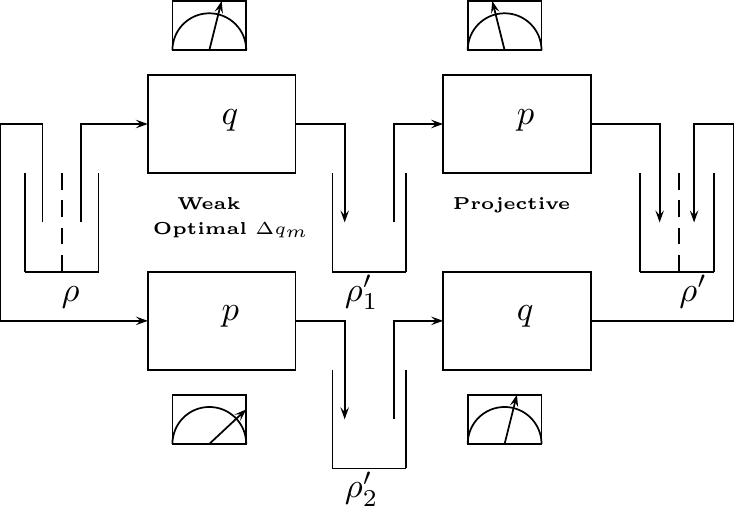}
\caption{The algorithm to implement our scheme
where we first divide the ensemble $\rho$ into two
parts. For one half of the ensemble we measure
position $q$ weakly (the weakness being defined by
the initial spread in position $\Delta q_{m_1}$ of the
meter) leading to a disturbed ensemble $\rho'_1$.
On every member of the ensemble $\rho'_1$ we carry out
a projective measurement of momentum $p$.  With
the other half of the initial ensemble $\rho$,
momentum $p$ is measured weakly leading to a
disturbed ensemble $\rho'_2$ on which  a
projective measurement of $q$ is carried out.
\label{flowchart}}
\end{figure}
In order to perform complete state tomography of
any Gaussian state of the form discussed earlier,
we are required to estimate the center of the
Gaussian Wigner function, $\left(q_0,p_0\right)$ and the
spreads $\Delta q$ and $\Delta p$. Hence it is
necessary to measure both position $q$ and momentum $p$
of the system as accurately as possible. To this
end, we divide the initial ensemble of identically
prepared systems into two equal parts.  On every
member of one part we perform a weak measurement
of position $q$. The strength of the measurement
is governed by the initial squeezing of the
position quadrature of the meter determining the 
initial variance $\Delta
q_{m_1}$ of the meter state.  The larger the value
of $\Delta q_{m_1}$, weaker is the  measurement
strength and vice versa.  The meter reading is
recorded in each case and the final states of all
the members are collected to generate a second
ensemble.  The members of this ensemble are now
used as the initial states of a second
measurement, which is a projective measurement of
momentum $p$. As before the meter readings of this
measurement are noted.  Now the process is
repeated with the members of the second part of
the initial ensemble where we first measure $p$
weakly, with the strength of the measurement
determined through $\Delta q_{m_2}$ and then carry
out a projective measurement of $q$. In all
further analysis and discussions we take $\Delta
q_{m_1}=\Delta q_{m_2}=\Delta q_m$. A summary of
the procedure is illustrated in
Figure~\ref{flowchart}. The entire algorithm is
repeated over many runs to rule out statistical
fluctuations. It is worth noting that the initial
squeezing of the relevant quadrature which
determines the strength of the measurement is a
tunable parameter in our hand.  Although we call
certain measurements ``weak'', we actually mean
that it is not too strong to be projective and not
too feeble to induce large errors to the
measurement outcomes. The main point is that the
measurements are weak enough and do not cause the
complete collapse of the state so that it can be
used for subsequent measurements.  The
expectation values obtained from the $q$ and $p$
measurements are used to estimate the values of
$\left(q_0,p_0\right)$, and the spreads $\Delta q$
and $\Delta p$.

Looking at the  Equations~(\ref{wigsymp1})
and~(\ref{wigsymp2}) reveals  that the
information about the system has flowed into the
meter. In fact the meter is now centered over
$(q_0,p_0)$ which is  the center of the initial
system  state.  We carry out simulations using the
meter reading probabilities given in
Equations~(\ref{probweak1}),~(\ref{probstrong2}),
~(\ref{probweak2}) and~(\ref{probstrong2}).  We
take different ensemble sizes of member numbers
$6$, $8$, $10$ and $20$ respectively with randomly generated
Gaussian states. Each virtual experiment is
repeated over 1000 runs. The quantities $q_0$ and
$p_0$ for a state are estimated by taking the mean
over the $q$ and $p$ measurements while $\Delta q$
and $\Delta p$ are estimated from the
corresponding variances. The order of measurement
of $q$ and $p$ is reversed for the second part of
the ensemble to rule out the possibility of
preferential treatment of any of the observables.

In the scheme involving projective measurements
only, we divide the original ensemble into two
parts and perform $q$ and $p$ measurements
independently on the individual members of these
parts. No sequential measurements are possible here
because of the wavefunction collapses after the
measurement.

The accuracy of the state estimate is measured via
the following distance measures 
\begin{eqnarray}
d_1&=&\left(q_0-q_0^{est}\right)^2+\left(p_0-p_0^{est}\right)^2
\nonumber \\
d_2&=&\left(\Delta q-\Delta
q^{est}\right)^2+\left(\Delta q-\Delta
p^{est}\right)^2 \label{measure} 
\end{eqnarray}
where $q_0^{est}$, $p_0^{est}$, $\Delta q^{est}$
and $\Delta p^{est}$ are the estimated values of
$q_0$, $p_0$, $\Delta q$ and $\Delta p$,
respectively. The parameter $d_1$
is a measure of how well our method is able to
estimate the center of the Gaussian and $d_2$
gives a measure of how well the spreads of the
Gaussian have been estimated. The two measures
$d_1$ and $d_2$ represent closeness in position
and width of the estimated Wigner distribution
from the original Wigner distribution respectively.
We can immediately see that the lower these distances,
the better the estimates. For a perfect estimate the
values should go to zero.
\subsection{Performance of the scheme} 
To study the average performance of our scheme for 
squeezed displaced thermal states,
we begin by numerically generating 100 Gaussian
states at a particular temperature, with randomly
chosen values of displacement and squeezing.  To
generate these states, the value of the parameter
$u$ in Equation~(\ref{squeezing}) is varied
between $-1$ and $+1$ according to a uniform
distribution. Similarly, the centers of the
Gaussians are also chosen randomly using uniform
distributions between $-3$ and $+3$ for both $q_0$
and $p_0$.

With each of these 100 random states, we
numerically carry out the prescription given in
subsection~\ref{pres} on a fixed number of
identical copies of the state determining the
ensemble size.  The simulation is carried out with
the help of the results obtained in
section~\ref{scheme}.  The distance measures $d_1$
and $d_2$ used to compare the efficacy of our
method with projective measurements are computed.
Each experiment involving one Gaussian state is
repeated 1000 times to reduce statistical
fluctuations. The process is carried out with
ensembles of sizes $20$, $10$, $8$ and $6$.  For a
given ensemble size, the results for each member
are averaged over 1000 runs and then the distance
measures are averaged over the 100 states.  We
show that there is a clear advantage of using our
scheme when the ensemble size is small. The test
is carried out for three different sets of
Gaussian states corresponding to three different
temperatures given by $\kappa=1$, $0.9$ and $0.8$,
respectively.

\begin{figure}[t]
\centering
\includegraphics[scale=1]{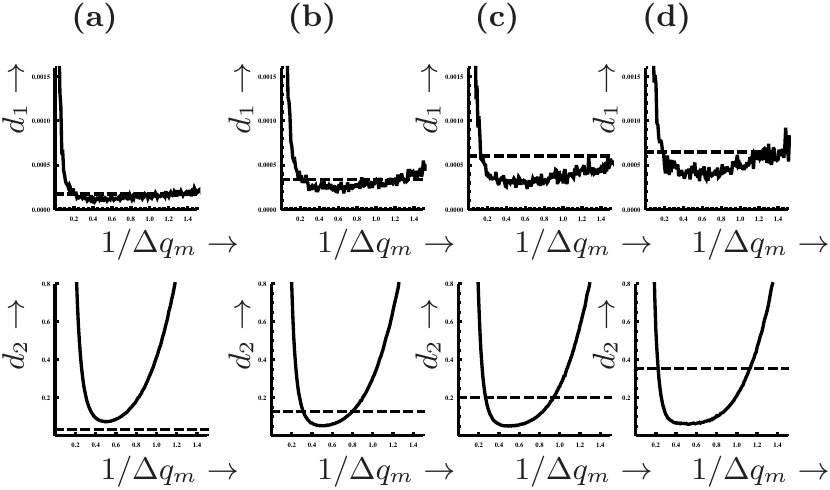}
\caption{The efficacy of our method as compared to
projective measurements for $\kappa=1$ using
averages over 100 random states further averaged
over 1000 runs. The behaviors of $d_1$ and $d_2$
are plotted with  $1/\Delta q_m$ for ensemble
sizes $(a)\; 20\; (b)\; 10 \;(c)\; 8$ and $(d)\;
6$. The corresponding projective measurement
results are plotted as dotted lines. While the
method performs well in estimating the position of
the Gaussian states for all ensemble sizes, as
represented by $d_1$, it provides a clear
advantage for estimating the spreads represented
by $d_2$ over projective measurements in the case
of a small ensemble of size $6$.  \label{K_1}}
\end{figure} 
\begin{figure}[t]
\centering
 \includegraphics[scale=1]{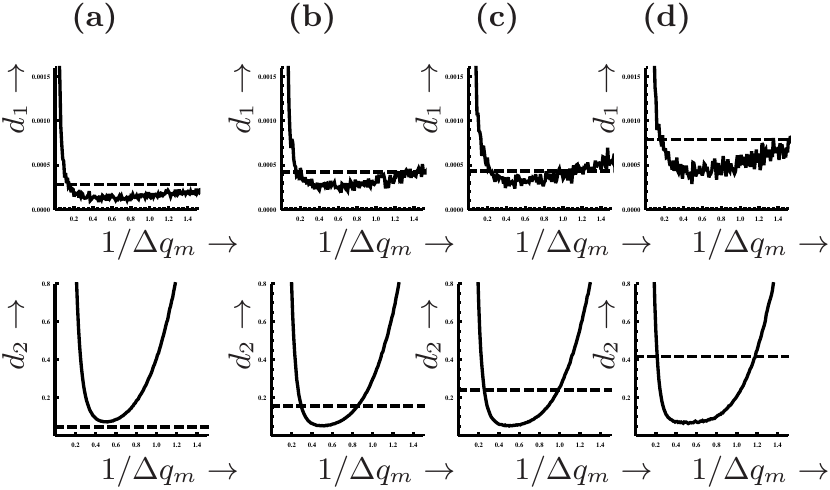}
 \caption{The efficacy of our method as compared
to projective measurements for $\kappa=0.9$ using
averages over 100 random states further averaged
over 1000 runs. The behaviors of $d_1$ and $d_2$
are plotted with  $1/\Delta q_m$ for ensemble
sizes $(a)\; 20\; (b)\; 10 \;(c)\; 8$ and $(d)\;
6$. The corresponding projective measurement
results are plotted as dotted lines. While the
method performs well in estimating the position of
the Gaussian states for all ensemble sizes, as
represented by $d_1$, it provides a clear
advantage for estimating the spreads represented
by $d_2$ over projective measurements in the case
of a small ensemble of size $6$.  \label{K_0.9}}
\end{figure} 
\begin{figure}[t]
\centering
\includegraphics[scale=1]{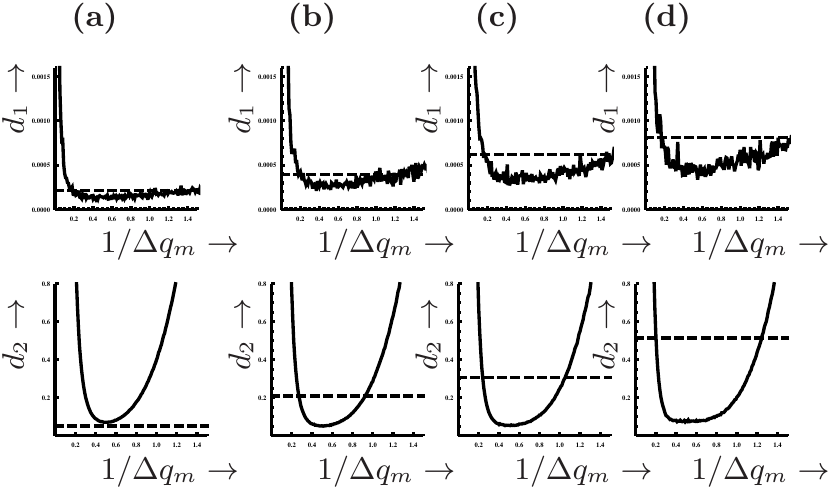}
\caption{The efficacy of our method as compared
to projective measurements for $\kappa=0.8$ using
averages over 100 random states further averaged
over 1000 runs. The behaviors of $d_1$ and $d_2$
are plotted with  $1/\Delta q_m$ for ensemble
sizes $(a)\; 20\; (b)\; 10 \;(c)\; 8$ and $(d)\;
6$. The corresponding projective measurement
results are plotted as dotted lines. While the
method performs well in estimating the position of
the Gaussian states for all ensemble sizes, as
represented by $d_1$, it provides a clear
advantage for estimating the spreads represented
by $d_2$ over projective measurements in the case
of a small ensemble of size $6$.  \label{K_0.8}}
\end{figure} 
The performance of state estimation of Gaussian
states via our weak measurement protocol is
compared to the corresponding performance of
projective measurements. This is done via
plots of the distance measures $d_1$ and $d_2$ vs
weakness parameter defined by the inverse of
squeezing $\Delta q_m$ of the meter state,
averaged over 100 such random states. The process
is carried out for four different small ensemble
sizes $20,\; 10 ,\; 8$ and $ 6$ and three
different absolute temperatures given by
$\kappa=1,\;\kappa=0.9$ and $\kappa=0.8$.

Let us first look at Figure~\ref{K_1} (a).  In
this case the distance measures $d_1$ and $d_2$
have been plotted with $\frac{1}{\Delta q_m}$ for
an absolute temperature given by $\kappa=1$ and
ensemble size $20$.  A low value of
$\frac{1}{\Delta q_m}$ indicates the meter
prepared as a wide Gaussian in the position space.
This corresponds to the weak measurement limit. A
very weak measurement introduces a large amount of
error in the measurement and this leads to a low
quality of state estimation. This can be seen from
the fact that the values of both $d_1$ and $d_2$,
on the left hand side of the plot for the weak
measurement method are much higher than those
involving only projective measurements
represented by the dotted line. Similarly, on the
right side of the plot, the meter is prepared as a
narrow Gaussian.  The corresponding measurement
limit for this side of the plot is that of strong
projective measurements.  Projective measurements
destroy the state of the system and hence using
the state for the second time leads to a low
quality of state estimation. Only for an
intermediate value of weakness, our method
performs better than projective measurements.
This is seen from the plot of $d_1$ going below
the dotted line representing the same distance
measure for the projective measurement. The plot
of $d_2$ attains its minimum for the intermediate
values of $\frac{1}{\Delta q_m}$ but remains above
the dotted line. It indicates that though our
method has worked in giving a better estimation of
the position of the Gaussian state, it does not
perform as well to provide an estimation of the
spreads of the Gaussian, in this particular case.

Figure~\ref{K_1} (b) shows the plot of the same
parameters for the same absolute temperature but
for a lower ensemble size of $10$. We find that
here our method proves to be more effective than
the projective measurements both for the
estimations of the position and the spread of the
Gaussian Wigner function. Moving on to
Figure~\ref{K_1}(c) and (d) which are for the
ensembles of sizes $8$ and $6$ we find that the
relative efficacy of the estimation for position
as well as the spread improves.

We repeat the same exercise with Gaussian states
with finite temperatures with
$\kappa=0.9$ and $\kappa=0.8$ as indicated in
Figures~\ref{K_0.9} and~\ref{K_0.8},
respectively. We observe the same trend as
observed 
for the zero temperature in all these
cases. Our method is not too effective in the
extremely weak or extremely strong regimes. It
works in the intermediate regimes depending upon
the size of the ensemble and its efficacy
increases with the lowering of the ensemble size.

In each of the plots, it is observed that the
distance measures attain small values for an
optimal value of squeezing. This is expected, as a
very large value of squeezing ushers in too many
errors into the ``weak measurement'', while a small
value causes a larger disturbance to the original
state.

We observe from Figures~\ref{K_1},~\ref{K_0.9}
and~\ref{K_0.8} that for an optimal range of
$1/\Delta q_m$ values, the weak distance measure
curves go below the projective measurement line
(represented by broken straight lines). In this
regime  of $1/\Delta q_m$ values, our method is
more effective than the projective measurement
state estimation.  The advantage is greater for
smaller ensemble size.  In fact for the
ensemble size of $20$ and $\kappa=1$, the
performances of the optimal weak measurement
method and projective measurement are almost equal
as can be seen in Figure~\ref{K_1}. However, as
the ensemble size decreases, a clear advantage
emerges for the proposed scheme.
There is no particular change in the
advantage of our scheme relative to
projective measurements, with  change of
temperature 
as is evident from plots with different
temperature parameter $\kappa$.
\section{Concluding remarks} 
\label{conc}
In this paper, we have described our work on the
estimation of Gaussian states by a method
employing weak or unsharp measurements.  We use
phase space methods and the language of Wigner
distributions for state estimation.  We compare
our results with state estimation based on
projective measurements and show how one can do
better in certain parameter regimes.  Recycling of
states, where one makes more than one measurement
on a single copy before discarding it and
tenability of the strength  of the weak
measurement are the two main ingredients of our
scheme.  The strength of the measurement is
directly related to the amount of squeezing in the
initial pointer state and can be tuned at will and
we optimize the performance of our scheme with
respect to this weakness parameter.  The efficacy
of the scheme is tested over a randomly chosen
subset of Gaussian states.  We demonstrate that
the weak measurement based scheme produces a
Wigner distribution which is much closer to the
original Wigner distribution as compared to the
scheme based on projective measurements, for small
ensemble sizes.  As the ensemble size increases,
the relative advantage of our scheme decreases, as
seen in the comparative results for varying
ensemble sizes. The behavior is repeated over the
range of temperatures we have considered.

While in this work we have dealt with Gaussian
states with the maximum spread along the $q$ or
$p$ axes it will be interesting to extend the
scheme to general Gaussian and non-Gaussian
states.  Another interesting direction that we
are following up is to compare our results
with schemes similar to the Arthurs and Kelly
setup where position and momentum are measured
together.
\begin{acknowledgements}
This work has been funded by the Department of Science and 
Technology (DST), India, under Grant No. DST-15-0079.
\end{acknowledgements}
%
\end{document}